      [1999/12/01 v1.4c Il Nuovo Cimento]
\documentclass{cimento}

\usepackage{graphicx}  
\title{Ultra High Energy Nuclei Propagation}
\author{R.~Aloisio\from{ins:lngs}}

\instlist{\inst{ins:lngs} INFN - Laboratori Nazionali Gran Sasso, Assergi Italy}

\PACSes{\PACSit{98.70 Sa}{13.85.Tp}}

\begin{document}

\maketitle

\begin{abstract}
We discuss the problem of ultra high energy nuclei propagation in astrophysical backgrounds. We present a new analytical computation scheme based on the hypothesis of continuos energy losses in a kinetic formulation of the particles propagation. This scheme enables the computation of the fluxes of ultra high energy nuclei as well as the fluxes of secondaries (nuclei and nucleons) produced by the process of photo-disintegration suffered by nuclei.
\end{abstract}

\section{Introduction}
A clear understanding of the composition features of the UHECR spectrum passes through a detailed study of the propagation of UHE protons and nuclei in astrophysical backgrounds. In particular, UHE nuclei interacting with such backgrounds suffer two main process: (i) photo-disintegration, giving rise to the production of secondary nucleons (hereafter protons) and lighter nuclei, (ii) pair production, that produces a depletion in the nucleus energy. The study of the photo-disintegration process of UHE nuclei has already attracted the attention of several authors that have studied the propagation of these particles with the aim of Monte Carlo (MC) simulations \cite{Stecker76},\cite{Epele98},\cite{Stecker99},\cite{Bertone02},\cite{Yamamoto04},\cite{Allard06} (for a detailed list of references see also \cite{NoiPRD}). One of the pioneering works in this field was performed in \cite{Stecker76}, where the authors introduced, for the first time, a very useful parameterization of the photo-disintegration cross-section, this result was up-dated in \cite{Stecker99}. Recently, a more detailed parameterization of the photo-disintegration cross-section was presented in \cite{Allard05}, this computation confirms the findings of \cite{Stecker76} and \cite{Stecker99} at energies below $10^{20}$ eV, showing some deviations at the extreme energies. The studies performed so far have shown the important role that the Infrared/Optical Background (IRB) plays in nuclei propagation, this is an important result that, on the other hand, complicates the understanding because of the less precise determination of such background respect to the well known Cosmic Microwave Background (CMB). 

In this paper we will briefly present a new analytical computation scheme that, based on a kinetic approach in the approximation of continuos energy losses, enables the computation of UHE nuclei propagation in any background, with the determination of the expected fluxes on earth of the primary injected particles as well as of all secondaries (nuclei and nucleons) produced in the photo-disintegration process. Following \cite{NoiPRD}, in order to explain the method proposed, we will restrict the discussion to the case of the CMB radiation alone. 

\section{Nuclei propagation and energy losses}
\label{losses}

Nuclei propagating in the intergalactic medium suffer a degradation of the Lorentz factor and are photo-disintegrated with a change in the propagating nuclei specie. In particular, the Universe expansion and the process of pair production on astrophysical backgrounds are responsible for the Lorentz factor depletion, with the conservation of the nuclei species (i.e. atomic mass number A). On the other hand, the interaction of nuclei with astrophysical backgrounds is also responsible for the photo-disintegration of nuclei, 
this process changes the nuclei species conserving the nucleus Lorentz factor. 
The process of pair production is efficient only on the CMB background field, while the 
photo-disintegration process depends also on the IRB. As discussed in \cite{NoiPRD}, once determined the energy losses, namely the two quantities $\beta_A=1/A (dA/dt)$ and $\beta_\Gamma=1/\Gamma (d\Gamma/dt)$, we can compute the survival history of nuclei during their journey from the source to the observer. 
Specifying the initial conditions at red-shift $z=0$, namely the observed nuclei specie ($A_{obs}$) at the observed energy $E_{obs}=A_{obs}\Gamma_{obs}m_N$, solving the evolution equations $(\beta_A,\beta_\Gamma)$ we can trace the nucleus back to the source where it was injected as $A_0$ with the generation energy $E_g=A_0 m_N \Gamma_g(\Gamma_{obs},A_{obs},A_0)$ (being $m_N$ the proton mass).   

\section{Fluxes}
\label{fluxes}

We consider an expanding universe homogeneously filled by the sources of the accelerated primary nuclei of a single fixed specie $A_0$, with a generation rate per unit of co-moving volume given by $Q_{A_0} (\Gamma,z)=\frac{\gamma_g-2}{m_N A_0} {\cal L}_0 \Gamma^{-\gamma_g}$, where $\gamma_g>2$ is the generation index and ${\cal L}_0$ is the source emissivity (i.e. the generated energy per unit of co-moving volume and per unit time) at $z=0$. The co-moving space density of nuclei $n_A(\Gamma_A,t)$ of a fixed specie $A$ (primary $A=A_0$ or secondary $A<A_0$) can be determined as the solution of the kinetic equation given by
\begin{equation}
\frac{\partial n_A(\Gamma_A,t)}{\partial t} - 
\frac{\partial }{\partial \Gamma_A}\left [ b_\Gamma(\Gamma_A,t) n_A(\Gamma_A,t) 
\right ] + \frac{n_A(\Gamma_A,t)}{\tau_A(\Gamma_A,t)} = Q_A(\Gamma_A,t),
\label{eq:kin}
\end{equation}
where $b_\Gamma=\Gamma_A \beta_\Gamma = (\beta_{\rm pair}+ \beta_{\rm ad})\Gamma_A $ is the rate of the Lorentz-factor loss, $Q_A$ is the injection of nuclei (primaries $A=A_0$ or secondaries $A<A_0$). The photo-disintegration process in (\ref{eq:kin}) is interpreted as a decaying process that simply depletes the flux of nuclei with an associated nucleus "lifetime" given by $\tau_A^{-1}=dA/dt$. As discussed in \cite{NoiPRD} the solution of the kinetic equation at red-shif zero $n_A(\Gamma)$ is given by 

\begin{figure}[!ht]
\begin{center}
\includegraphics[width=0.49\textwidth]{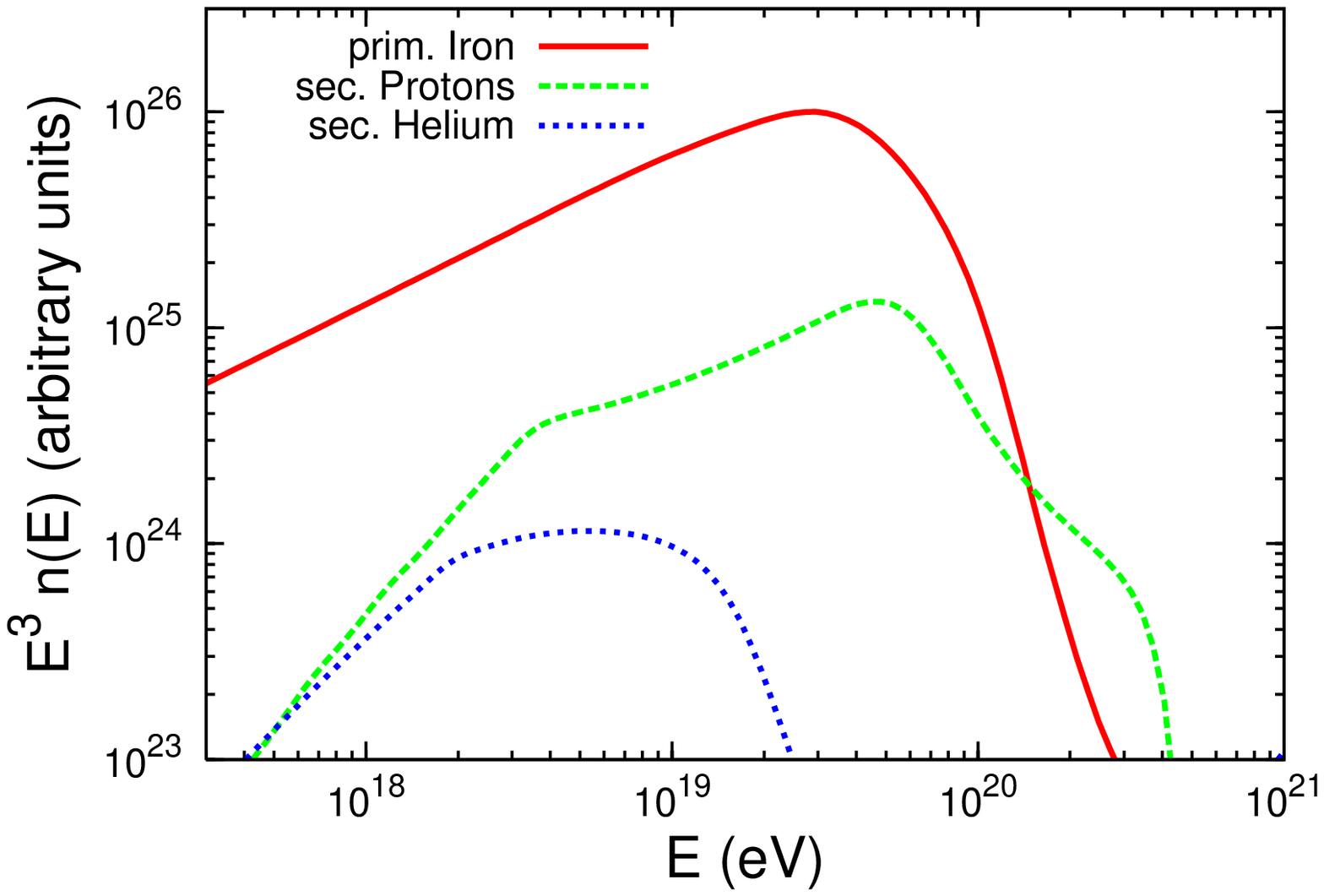}
\includegraphics[width=0.49\textwidth]{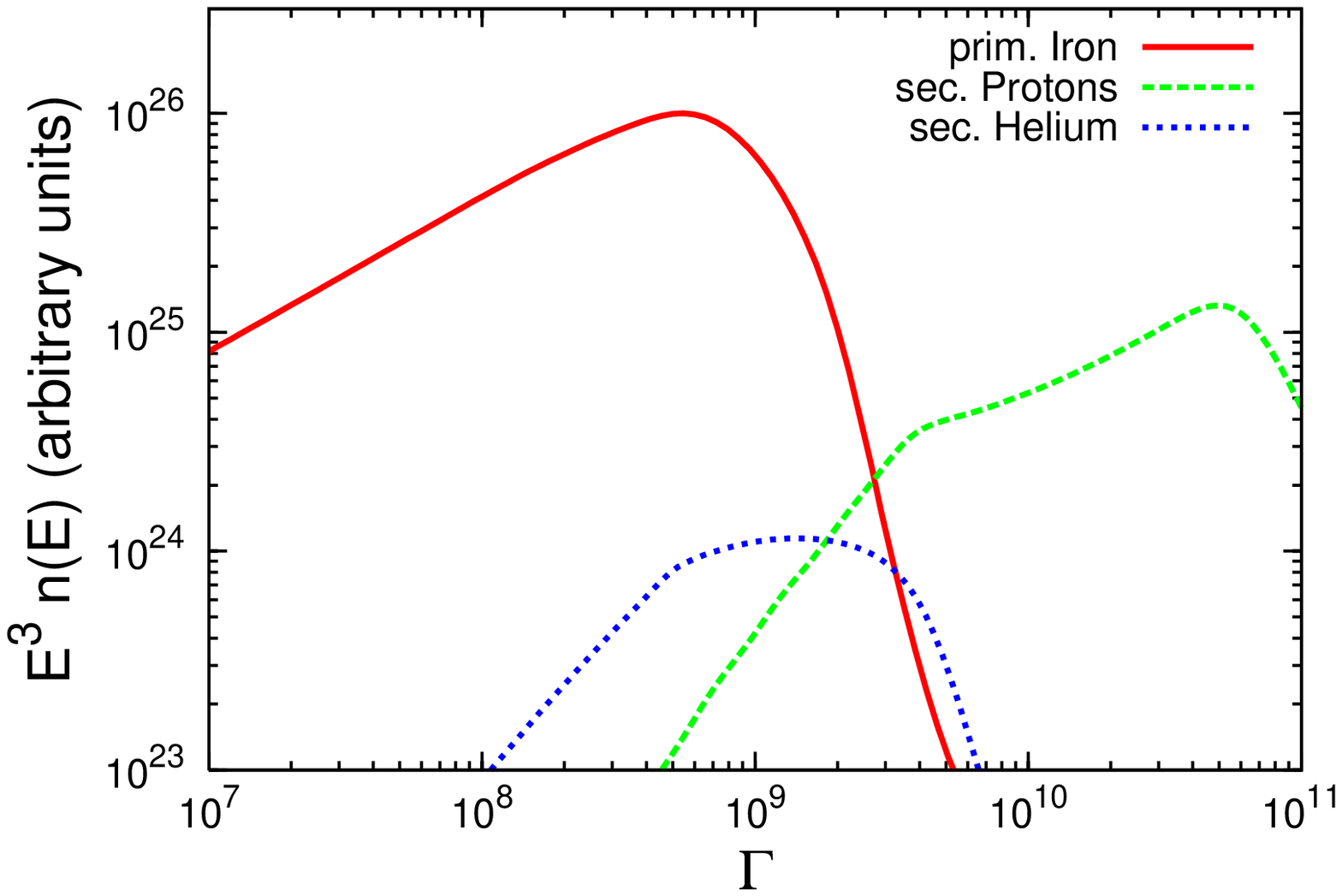}
\caption{ Fluxes of primary Iron secondary protons and Helium (with $\gamma_g=2.3$ and $E_{max}^{Fe}=Z_{Fe} 10^{21}$ eV) as function of the energy (left panel) and of the Lorentz factor (right panel). Only the CMB background filed is taken into account.}
\label{fig}
\end{center}
\end{figure} 

\begin{equation}
n_A(\Gamma)=\int_{z_{\rm min}}^{z_{\rm max}} dz \left | \frac{dt}{dz} 
\right | Q_A(\Gamma_A,z) 
\frac{d\Gamma_A}{d\Gamma} e^{-\eta(\Gamma_A,z)}
\label{eq:fluxA}
\end{equation}
where $z$ is the redshift of 
$A$-nuclei production, $\Gamma_A={\mathcal G}(A,\Gamma,z_0,z)$ is the Lorentz factor of the nucleus $A$ obtained by evolving $A$ from $z_0=0$ (when  Lorentz factor is $\Gamma$) up to z. The quantity $\eta(\Gamma_A,z)$ takes into account the 
photo-disintegration lifetime of the nucleus $A$, and it is given by 
$\eta_A=\int_0^z dz' (dA/dz')$. As discussed in \cite{NoiPRD}, the explicit expression of $Q_A$ is determined imposing the conservation of the number of particles along the evolution trajectory of the nucleus $A$ (with $\Gamma$ at $z=0$) and assuming that $A$ is always produced by the photo-disintegration process suffered by its father $(A+1)$. Under these assumptions one has: $Q_A\ne 0$ only inside the interval $(z_{min},z_{max})$, as given in \cite{NoiPRD}, with $Q_A(\Gamma_A,z)=Q_{A_0}(\Gamma_g,z_g) \frac{1+z}{1+z_g} \frac{d\Gamma_g}{d\Gamma_A}$, being $z$ the red-shift of $A$ production, $z_g$ the red-shift of the injected primary $A_0$ and $d\Gamma_g/d\Gamma_A$ as in \cite{NoiPRD}. In the present paper we take into account the CMB radiation field alone, under this assumption the flux of secondary protons can be fairly computed assuming an instantaneous photo-disintegration of the primary nucleus $A_0$  \cite{NoiPRD}, giving rise to an instantaneous production of $A_0$ nucleons. Once determined the injection of protons $Q_p=A_0 Q_0$ their flux is simply determined as in \cite{NoiPRD}. 
In figure \ref{fig}, assuming $A_0=56$, we show the flux of primary Iron, secondary protons and secondary Helium in the case of $\gamma_g=2.3$ with a maximum energy of acceleration $E^{Fe}_{max}=Z_{Fe} 10^{21}$ eV. 

\acknowledgments
I'm grateful to V. Berezinsky, A. Gazizov and S. Grigorieva with whom the present work was done. This paper was partially funded by the Italian Space Agency under the contract ASI-INAF I/088/06/0 for theoretical studies in High Energy Astrophysics.

\end{document}